# Control-Scheduling Codesign: A Perspective on Integrating Control and Computing

Feng Xia and Youxian Sun
*State Key Laboratory of Industrial Control Technology, Zhejiang University*
Hangzhou 310027, China
Email: f.xia@ieee.org

*Abstract* – Despite rapid evolution, embedded computing systems increasingly feature resource constraints and workload uncertainties. To achieve much better system performance in unpredictable environments than traditional design approaches, a novel methodology, control-scheduling codesign, is emerging in the context of integrating feedback control and real-time computing. The aim of this work is to provide a better understanding of this emerging methodology and to spark new interests and developments in both the control and computer science communities. The state of the art of control-scheduling codesign is captured. Relevant research efforts in the literature are discussed under two categories, i.e., control of computing systems and codesign for control systems. Critical open research issues on integrating control and computing are also outlined.

## I. INTRODUCTION

With the widespread emphasis on pervasive and ubiquitous computing technologies, the use of embedded systems within the engineering community has increased dramatically since some years ago, and this trend is expected to continue in the near future [1]. It has been recognized that over 99% of all microprocessors are now used for embedded systems that control physical processes and devices in real-time [2]. In contrast to general-purpose desktop systems, most of embedded computing platforms are typically resource limited [3,4], due to, e.g., cost constraints related to mass production and increasing industrial competition. On the other hand, in many fields the requirements on application functionality of embedded computing systems increase substantially. The systems are becoming over more complex. Consequently, dedicated processors are no longer available for most tasks, which have to share the same processors with each other. This could potentially cause the system workload to be highly varying.

In such systems, especially where the operating environments change over time or system reconfigurations are needed, contention for computing resources has become a central concern since the uncertain resource availability can result in performance problems [3-5]. As a consequence, the challenge in designing embedded computing systems upon such platforms turns to be how to implement applications that can execute efficiently on limited resources, while meeting application requirements such as performance, timeliness, flexibility, and so on.

It is well-known, especially in the computer community, that real-time task scheduling is a key lever in computing systems for system performance and resource usage. From 1970s, a large number of results have been reported in the literature of real-time scheduling theory [6]. Despite this, classical real-time scheduling algorithms, such as Rate Monotonic (RM) and Earliest Deadline First (EDF), are built upon complete knowledge about execution time, deadline, etc. of the task set. Often it is assumed that the timing constraints of a real-time task are precisely known *a priori*. In practical applications, however, this hypothesis is usually unrealistic. For instance, a task usually has a variable execution time that is neither known nor observable, and obtaining the worst case execution time (WCET) of a task is always very hard, if not impossible. From the control point of view, these algorithms are all open-loop [7]. Typically, the resources are allocated in a static (pre-specified) fashion. Once established at system set-up, schedules are not dynamically adjusted based on continuous feedback. Although these scheduling algorithms can perform well in resource sufficient environments, they are prone to cause a highly under-utilized system due to, e.g., pessimistic estimations of workload or WCET based design. Moreover, their performance degrades rapidly in resource insufficient environments. Particularly, in unpredictable environments where the workloads cannot be accurately modelled, as well as in systems built upon commercial-off-the-shelf (COTS) components, they will perform very poorly.

The severe and possibly variable constraints in computing resources as well as the poor performance of existing task scheduling algorithms call for a fundamentally different approach to resource scheduling. It has been recognized that feedback control is an important technology that can be employed in the design of embedded computing systems [6]. Over the years, a plenty of theory and methodology have been well-established and mathematically well-founded in the field of feedback control, which is ideal for handling uncertainties. Introducing control mechanisms into computing systems will allow the system to adapt to changes in the environment, the workload, or even to changes in the system architecture due to reconfigurations or failures. The result could be improved system performance and higher resource utilization.

While feedback control can serve as a scientific underpinning for computing systems, control systems themselves constitute an important subclass of embedded computing systems [3]. With increased complexity, it is not uncommon that several control tasks have to compete for one embedded processor. Therefore, the overall control performance will not only depend on the design of control algorithms, but also rely on the efficient scheduling of the shared computing resources. Unfortunately, the design of embedded control systems is often based on the principle of *separation of concerns* [8]. This separation is based on the assumption that feedback controllers can be modelled and implemented as periodic tasks that have a fixed period, a known WCET, and a hard deadline, which has also been widely adopted by the control community for developing sampled control theory. Although this assumption allow the control community to focus on its own problem domain without worrying about how task scheduling is done and they release the scheduling community from the need to understand how scheduling impacts control performance, the control task does not always utilize the available computing resources in an optimal way, and the assumptions of the simple task model are also overly restrictive with respect to the characteristics of many control loops. For instance, many control loop deadlines are not always hard. Instead most practical control systems can tolerate occasional deadline misses. As a result, the resulting quality of control (QoC) of real-time control systems that are designed based on this separation of concerns of control and scheduling would be worse than possible, and in extreme cases unacceptable with instability.

In order to cope with the resource constraints in embedded systems, codesign is needed at different levels, e.g. hardware/software codesign and the codesign of the mechanical design and the electrical design. In this paper, we will explore an emerging field in terms of *control-scheduling codesign*, which is motivated by the above requirements for integrating feedback control and real-time computing from both the control and computer science perspectives. A special focus is on the integration of feedback control and real-time scheduling in the context of embedded computing systems. Our aim is to provide an overview of current efforts in this field and to highlight the relevant research issues, thus sparking new interest and development in both the control community and the real-time computing community.

The rest of this paper is structured as follows. Section II describes the codesign problem of control and scheduling, while specifying the basic idea behind this methodology. The state-of-the-art work related to control/scheduling codesign is classified into two categories, feedback control of computing systems and codesign for real-time control. Important approaches in both categories are discussed in Section III and IV respectively. In Section V, we point out the most critical open research issues in this field. Finally, we give concluding remarks in Section VI.

## II. THE CODESIGN OF CONTROL AND SCHEDULING

Feedback control theory has a long history, and represents a well-developed analytic foundation for performance control in physical systems. Traditionally, the uncertainties to be attacked are associated with the physical plant to be controlled. However, as a matter of fact, the theory and design principles can also be applied to arbitrary systems containing uncertainties, e.g., real-time computing systems with uncertainties in, e.g., workload and resource utilization patterns [3,6]. Control theoretical approaches could be used to model, analyze, and design embedded computing systems. Furthermore, feedback control brings significant advantages to real-time computing systems. It is intuitive that a system designed using feedback control theory and methods could be quite robust towards external and/or internal disturbances and uncertainties. Therefore, control techniques can be properly employed to compensate for or eliminate the negative impact of certain implementation platforms. In this way, the flexibility of the system could be enhanced. In particular, when feedback control is used in conjunction with real-time scheduling, precise schedulability models are no longer needed. Instead, the system can adjust its resource allocation dynamically to achieve the desired temporal behaviour.

As have been mentioned, traditional control systems design methodology cannot achieve the optimal possible QoC for real-time control systems that feature resource constraints and dynamic workloads. For these types of control applications, a codesign approach would be very advantageous. With control-scheduling codesign in the context of real-time control, it implies that the control system is designed by taking both the control and computing aspects into account simultaneously. In other words, the controller algorithm design stage and the system implementation stage, which are separated traditionally, would be integrated. The use of codesign in control systems has the potential to improve the control performance. From the standpoint of resource utilization, it is also in most cases desirable to allocate computing resources to different tasks such that the resulting overall system performance is optimized. The codesign of control and scheduling is exactly a promising methodology to achieve this objective.

One can find an initial yet extensive introduction, mainly from the control perspective, of control/scheduling codesign in [8]. To this day, there still lacks of a general formulation of the codesign problem of control and scheduling. From a control viewpoint, Årzén *et al.* [3] informally state the control and scheduling codesign problem in the uniprocessor case as follows: *Given a set of processes to be controlled and a computer with limited computational resources, design a set of controllers and schedule them as real-time tasks such that the overall control performance is optimized*. Actually, the control and scheduling codesign is applicable to not only embedded control systems but also other multitasking embedded computing systems where real-time scheduling is involved.

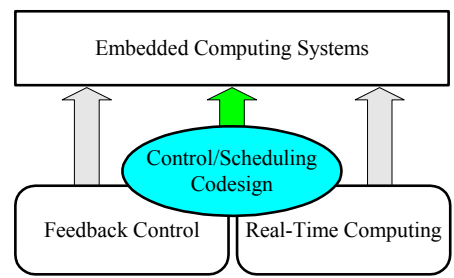

**Fig.1.** Control/scheduling codesign as a result of the integration of feedback control and real-time computing

Generally speaking, the basic idea behind control and scheduling codesign is to combine real-time scheduling theory and feedback control theory in embedded computing systems design, so that the available computing resource is optimally utilized and the overall system performance is maximized. As regarding an embedded web server system, for instance, the objective of codesign will be to achieve QoS (Quality of Service) guarantees on service delay. As illustrated in Fig. 1, two major areas, i.e., feedback control and real-time computing, are involved in the codesign of control and scheduling. They altogether provide the major theoretical and technical supports for designing embedded computing systems. A control-scheduling codesign approach can be either online or offline. While offline codesign is often performed with the goal of optimizing the use of the available resources, online codesign approaches usually just target system flexibility in resource management because obtaining optimal solutions would be considerably computationally-intensive, which results in overly large overhead.

In recent several years, an increasing number of research efforts from both the computing side and the control side have been made in this emerging field. According to the ultimate goals of the systems considered, most existing relevant work could be classified into two categories. One is feedback control of computing systems where the methodology of feedback control real-time scheduling is explored in most cases. Usually, feedback control theory is used as a scientific underpinning to improve the real-time performance of computing systems. The other is codesign for real-time control systems where the major purpose is to maximize the QoC as much as possible. In the next two sections, we will discuss critical issues and existing efforts in these two categories respectively.

### III. CONTROL OF COMPUTING SYSTEMS

Applying feedback control theory in real-time computing systems is an area that currently is receiving a lot of attention. This approach is sometimes called *feedback (control) scheduling* [3,6-8]. The basic idea behind is to treat the scheduling problem as a feedback control problem. A feedback control loop is introduced into the resource management in computing systems. Regarding the real-time computing systems as controlled dynamics with defined error terms, feedback schedulers are designed using feedback control theory. The objective of feedback scheduling is to increase the flexibility with respect to uncertainties in resource utilization. Instead of pre-allocating resources based on offline analysis the resources are allocated dynamically online, based on feedback from the actual resource utilization. In principle, feedback control theory can be applied to the allocation of any type of resources in real-time computing systems.

Important efforts in the application of control theory to operating systems, particularly from the real-time computing community, include [7,9-21]. In [9,10], a general architecture for feedback control real-time scheduling and a new real-time scheduling algorithm called Feedback Control Earliest Deadline First (FC-EDF) is presented. A PID (Proportional-Integral-Derivative) controller regulates the deadline miss-ratio for a set of soft real-time tasks with varying execution times, by adjusting their CPU utilization. The new scheduling algorithm has proved to be robust in overload situations. In [11] the approach is extended. The resulting hybrid controller scheme, named FC-EDF$^2$, gives good performance both during steady state and under transient conditions. The framework is further generalized in [7]. Abeni *et al.* [12] apply control theory to a reservation-based feedback scheduler and provide a precise mathematical model of the scheduler. This approach has also been implemented in a Linux environment [13]. In [14], feedback is used in combination with elastic scheduling to estimate the actual load and adapt task periods to reach a desired utilization factor. The scheme in [15] allocates to each thread a percentage of CPU cycles over a period of time, and uses a feedback-based adaptive scheduler to assign automatically both proportion and period. For distributed real-time systems, the authors of [16] propose a double-loop feedback scheduler, whose objective is to keep the deadline miss ratio near the desired value and achieve high CPU utilization. With the similar objective, Sahoo *et al.* [17] design PI (Proportional-Integral) controller and H∞ controller for closed loop scheduling.

Since almost all real-world systems are essentially nonlinear, stochastic, time-varying, simple approaches such as PID control do not always work well. To attack this problem, several techniques from the control community have been employed in resource scheduling within operating systems. For example, aiming at the time-varying and non-linear characteristics of real-time CPU scheduling model, a soft real-time scheduling algorithm based on the hybrid adaptive feedback control architecture is presented in [18]. Abeni *et al.* [19] propose a closed-loop method for on-line adapting the fraction of assigned resource to the task requirements. The approach is based on adaptive control techniques and has resulted to be effective in a significant set of real-life experiments. In [20], Lawrence *et al.* present an approach to adaptive CPU scheduling that uses optimal feedback control to achieve the best allocation of CPU shares to periodic tasks. As an effective method to enhance the performance of feedback scheduling approaches, Amirijoo *et al.* [21] quantize the disturbance present in the measured variable as a function of the sampling period and propose a measurement disturbance suppressive control structure.

Another important real-time computing area into which control theory has been applied is high-performance server systems. Some challenges in the development of closed-loop systems for server systems are discussed in [5]. There are three types of control problems that are typically addressed. The first is to provide a capability for enforcing service level agreements in that customers receive the service levels for which they contracted, e.g. [22]. A second problem is to regulate resource utilizations so that they are not excessive, e.g. a mixture of queueing and control theory used to regulate the Apache HTTP Server [23]. The third problem that is often addressed is to optimize the system configuration, such as to minimize response times [24].

Recently, researchers begin to exploit feedback control methodology in dynamic voltage scaling (DVS) for power aware computing systems, with the goal of reducing energy consumption while guaranteeing required system performance. Lu *et al.* [25] describe formal feedback control algorithms for DVS in multimedia systems. The goal is to reduce multimedia decode power while maintaining a desired playback rate. A Feedback-DVS framework for hard real-time systems is proposed by Zhu and Mueller [26]. By combining feedback control with DVS schemes, it produces energy-efficient schedulers for both static and dynamic workloads. Kandasamy *et al.* [27] present a model predictive control framework to minimize the power consumed by a processor while satisfying QoS requirements of a varying workload. Following the methodology of feedback scheduling, Xia *et al.* [28] present a control theoretical DVS that facilitates tradeoffs between energy consumption and control performance through controlling the CPU utilization at a considerably high level.

Successful applications of feedback scheduling methodology can be found today in many other areas such as ORB middleware [29], autonomous vehicle systems [30] and networked control systems [31]. Resource management middleware frameworks such as ControlWare [32] and AutoTune Agents [33] are developed. A book on the application of control theoretical approaches for computing systems has also appeared [34].

## IV. CODESIGN FOR CONTROL SYSTEMS

In this section, we capture the second category of research efforts on control/scheduling codesign, where the focus is on real-time control. The computing system considered here usually consists of a set of digital control loops. Each controller is realized as a separate period task. Consequently, the main computing resource of concern in these systems is the CPU time [3]. The primary goal of codesign approaches becomes optimizing QoC under CPU resource constraints. In the literature, there are both offline and online approaches to control/scheduling codesign of control systems.

### A. Offline Codesign

It is intuitive that the ability to make an integrated offline design of control algorithms and scheduling algorithms is a prerequisite for online codesign of control and scheduling [8]. Offline codesign of control and scheduling allows an incorporation of the availability of computing resources into the control systems design by utilizing the results of real-time scheduling theory. The optimization of control performance subject to schedulability has firstly been treated in [35], where an algorithm is proposed to translate a system performance index into task sampling periods, considering schedulability among tasks running with preemptive priority scheduling. In [36] and [37], algorithms are proposed that optimizes the expected control performance under the constraint that the task set is schedulable. An approach to optimization of sampling period and input-output latency subject to performance specifications and schedulability constraints is presented in [38] and [39]. An integrated real-time control design approach is also presented in [40]. In [41], sampling periods for a set of controllers are chosen such that a certain robustness measure is maximized. The integration of static cyclic scheduling and optimal (LQ) control is the topic of [42]. The proposed solution contains the periodic task schedule as well as the state feedback gains of the controllers.

An alternative approach to offline codesign is eliminating the jitter within control loops, which greatly simplifies the performance optimization problem. Focusing on the jitter problem itself, some works have presented specific scheduling-based solutions, e.g. [43-47]. Crespo *et al.* [43] propose a method to determine the minimum interval where the control action or the data acquisition has to be allocated avoiding the jitter effects on control tasks. Balbastre *et al.* [44] extend the previous results providing also new schedulability analysis. Albertos *et al.* [45] suggest a method to reduce output jitter variability and its degrading effects on control performance by splitting control tasks with a new priority assignment. In a similar way, Cervin [46] shows that by scheduling two parts of a control algorithm as separate tasks the computational delay can be significantly reduced and thus the system performance is improved. In [47], the control server, a real-time scheduling mechanism tailored to control and signal processing applications, is presented.

There are also two controller design approaches to the problem of scheduling-induced jitter. The basic idea is to make controllers tolerate or account for scheduling effects. The first approach is to design the controller to be robust against jitter. The second approach is to actively compensate for the timing variations in each period. Many examples of these approaches can be found in the control community, e.g. [48].

### B. Online Codesign

Although offline methods can handle application requirements, they cannot easily deal with timing uncertainties due to, e.g., workload variations [1]. Thus it can be useful to consider more dynamic solutions, i.e. online codesign of feedback control and real-time scheduling. In order to provide QoC guarantee even in resource constrained dynamic environments, the execution of control tasks should be adapted with flexible scheduling of available computing resources, so that online tradeoffs between control performance and computing resource requirements are achieved.

Different methods for runtime tradeoff between control performance and schedulability have been proposed in [49-52]. Online adjustment of sampling periods in order to avoid CPU overloads is the topic of [49]. A number of different rate modulation algorithms suitable for RM scheduling are given. A more control-theoretic approach to rate modulation is taken in [50], where each controller is associated with a cost function. In [51] a QoS renegotiation scheme is proposed as a way to allow graceful degradation in cases of overload, failures or violation of pre-runtime assumptions. In [52], an integration of load driven online scheduling with direct digital design to optimize control performance as a function of varying workload is presented.

A special interest in this area is on the use of feedback scheduling methodology in multitasking control applications. As Fig.2 shows, the basic idea is to construct an outer feedback loop in addition to traditional process control loops. It is used to control the scheduling parameters, e.g., the CPU utilization, based on QoC requirements. Typically, the feedback scheduler controls the CPU activity according to the computing resource availability and workload variations by dynamically adjusting the timing attributes of controller tasks.

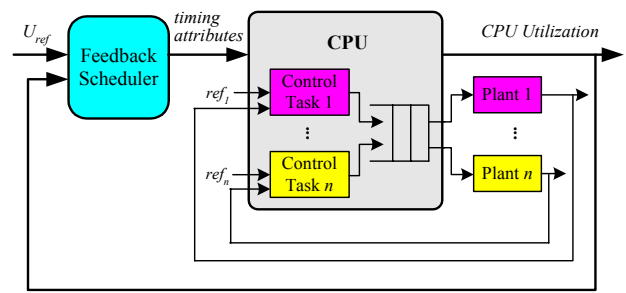

**Fig.2.** Feedback scheduling of control systems

Cervin and Eker [53] present a feedback scheduling mechanism for hybrid controllers where the execution time may change abruptly between different modes. The proposed solution attempts to keep the CPU utilization at a high level, avoid overload, and distribute the computing resources evenly among the tasks. Eker *et al.* [54] design an optimal feedback scheduler to distribute computing resources over a set of real-time control loops in order to optimize the total control performance. Its approximation versions are exploited in [55], where feedback scheduling is performed by simple rescaling of the task periods.

In the framework of embedded control, some researchers attempt to construct the feedback scheduler using modern control technologies. In [56], Xia *et al.* utilize neural network technique in feedback scheduling of a set of control tasks, and provide a fast, almost optimal solution. In [57,58], Xia *et al.* suggest an intelligent control theoretic approach to feedback scheduling based on fuzzy logic control technology, with the aim of providing flexible QoC management in the presence of timing uncertainty and imprecision. Robust and adaptive solutions for real-time scheduling and control co-design are dealt with in [59-61]. A delay-dependent feedback scheduler has been designed for control systems in [59]. It regulates the resource utilization according to the estimated execution times. The actuators are the tasks periods and a H∞ control approach provides robustness with respect to modelling errors. In [60] a processor load regulation has been presented based on a simple scheduling model and H∞ synthesis. The application of robust codesign to robot control is conducted in [61].

Feedback scheduling of anytime controllers is the topic of [62-64], where the execution time instead of period of each task is dynamically adjusted. Preliminary results on dynamic scheduling of model predictive controllers in which a quadratic optimization problem is solved iteratively in every sample are presented in [62]. A feedback scheduling approach is employed in [63] to attack the impact of dynamic resource constraints on a class of iterative control algorithms. In [64], a fuzzy feedback scheduler is proposed to improve the performance of iterative optimal control applications with imprecise timing attributes.

An interesting alternative to task rescaling is given in [14], where an elastic task model for control tasks is presented. The relative sensitivity of tasks is expressed in terms of elasticity coefficients. Based on this model, periodic tasks can intentionally change their periods at runtime to provide a different QoS, and other tasks can automatically adapt their periods to keep the system underloaded. Online codesign of control and scheduling based on the elastic model has been reported in [65,66].

Also another notable work in this field is the QoC scheduling approach that is formulated in [67]. It offers the possibility of taking scheduling decisions based on the control information for each control task invocation, rather than using fixed timing constraints with constant periods and deadlines. In this context, Marti *et al.* [68] present an optimal resource allocation policy that maximizes control performance within the available resources. It is shown that by using feedback to dynamically allocate resources to controllers as a function of the current state of their controlled systems, control performance can be significantly improved. Velasco *et al.* [69] provide guidelines on how to establish codesign approaches aimed at improving control performance or coping with overload conditions.

Tradeoffs between control performance and CPU energy consumption in real-time control systems have been explored in the context of integrating feedback control and power-aware computing. Lee and Kim [70] propose a static solution to obtain optimal processor speed that minimizes the CPU energy consumption as well as a dynamic solution to overcome unavoidable deficiencies of the static solution and to further reduce the energy consumption of the overall system. Xia and Sun [71] present an enhanced dynamic voltage scaling (EDVS) scheme, where the methodology of direct feedback scheduling is employed. The primary goal is to further reduce energy consumption over pure DVS methods while satisfying QoC requirements in real-time control systems. The control theoretical DVS by Xia *et al.* [28] also targets multitasking embedded control systems.

## V. OPEN RESEARCH ISSUES

We can see from the above that a considerable amount of progress in the field of integrating feedback control and real-time computing have been made in recent years. However, in order to fully exploit control/scheduling codesign in computing systems design, a majority of research issues are still open and candidates for future research. In this section, some of them, both in theory and practice aspects, are discussed.

*1) Modelling embedded computing systems.* While the application of feedback scheduling in computing resources has proved beneficial, it also offers several important challenges for future research [3,5,6]. The most fundamental issue is modelling of priority-based systems for control purpose, which seems to be a prerequisite for the integration of control theory in real-time scheduling. Because the behavior of embedded computing systems typically does not obey any first principles, the models used in the design of feedback schedulers, which can be viewed as controllers from the viewpoint of feedback control, are in most cases generated from input-output measurements. Guidelines such as those on how to choose the most appropriate types of models for computing systems should be constructed. Should the target system be modelled in discrete time or continuous time? How to obtain a model with guaranteed precision? Systematic methodologies are needed for modelling computing systems that are inherently characterized by nonlinearities and uncertainties.

*2) Advanced control based feedback scheduling.* To this day, many basic control techniques such as PID, optimal feedback control and robust control have been employed in the construction of feedback schedulers. Attempts to design more effective feedback schedulers based on advanced control theories such as neural networks, fuzzy control and adaptive control also begin to appear in these years. However, a lot of problems, especially those related to timing uncertainty of task execution in resource insufficient embedded environments, remain unsolved. On the other hand, the control community has provided plenty of well-established theory that can be used to effectively handling these problems. How to use existing advanced control theories to model, design and analyze feedback schedulers is a challenging topic. While novel structures may be developed, the scheduling overhead is an additional issue to which one should pay adequate attention when exploiting relatively complex control algorithms in the framework of feedback scheduling.

*3) Control theoretical issues in codesign.* In computing resource constrained control systems, adaptive scheduling schemes could be explored to dynamically distribute resources among real-time tasks with respect to resource availability variations. However, timing attribute changes and jitters resulting from dynamic scheduling strategies naturally lead to many interesting control-theoretical issues [3]. For example, a lot of adaptive scheduling schemes involve dynamically adjusting the sampling periods of control tasks during run time. This directly incurs sampling period jitters within control loops, and hence inversely degrades the QoC. Intuitively, as the changes of sampling periods become more frequent, the negative impact of these jitters will be more significant. Theory and methods must be developed to compensate for sampling period jitters. In addition, time-varying, possibly uncertain control delays should also be dealt with. Ideally, a controller should be designed to be robust against both sampling period jitter and input-output jitter.

*4) Supporting tools development.* To facilitate the codesign of control and scheduling, it is necessary to have supporting tools for simulation, analysis, and synthesis of embedded computing systems. During the last years, such tools that allow co-simulation of control and real-time scheduling, e.g., Jitterbug, TrueTime, RTSIM, and Ptolemy II, have begun to emerge [72], from both the control and the computing communities. Despite these projects on codesign tools development, none of the available tools is mature enough to support the whole life cycle of codesign of control and scheduling in embedded systems. While research interests on control/scheduling codesign expand dramatically, almost all available tools are limited, either in control or real-time aspect. The key reason behind is the historically long-time separation between the control and computing communities. For example, Jitterbug is only applicable to linear systems, while TrueTime needs to improve execution time estimation. With novel theoretic approaches to integrating control and computing appearing one by one, more efforts are required to enhance the capability of existing simulation tools or develop definitely new ones based on a codesign viewpoint.

*5) Practical implementation of feedback scheduling.* Feedback scheduling has become an important methodology in dynamic codesign of control and scheduling. With different structures and algorithms, it enables better use of the computing resources and leads to better system performance. Although some applications of feedback scheduling have been examined with the help of simulation tools, little work focuses on their practical implementations, particularly in the control community. Implementation of feedback scheduling-based control systems is an almost completely open issue.

## VI. CONCLUSIONS

The various requirements of modern embedded computing systems from both the control and computing perspectives have increased the emphasis on codesign of control and scheduling. This emerging methodology has inspired a lot of efforts towards integrated control and computing. The use of feedback control techniques has been gaining importance in real-time scheduling as a means to provide predictable performance in the face of timing uncertainties. The contributions of efficient resource scheduling schemes to QoC improvement have also been gradually recognized. However, control/scheduling co-design is a fairly new area with a young life of only several years. We encourage more insight into the integration of feedback control and real-time computing and more development in solutions to the open research issues as described in this paper.

ACKNOWLEDGMENT

This work is partially supported by Key Technologies R&D Program of Zhejiang Province (No. 2005C21087), Academician Foundation of Zhejiang Province (No. 2005A1001-13), and Specialized Research Fund for the Doctoral Program of Higher Education (No.20050335020).